\def\fileversion{v2.5}
\def\filedate{4 October 1993}

\newdimen\@bls                    % \@b(ase)l(ine)s(kip)
\@bls=\baselineskip               % \@bls ~= \baselineskip for \normalsize
\advance\@bls -1ex                % (fudge term)
\newdimen\@eps                    %
\def\section{\@startsection{section}{1}{\z@}
  {1.5\@bls plus 0.5\@bls}{1\@bls}{\normalsize\bf}}
\def\subsection{\@startsection{subsection}{2}{\z@}
  {1\@bls plus 0.25\@bls}{\@eps}{\normalsize\bf}}
\def\subsubsection{\@startsection{subsubsection}{3}{\z@}
  {1\@bls plus 0.25\@bls}{\@eps}{\normalsize\bf}}
\def\paragraph{\@startsection{paragraph}{4}{\parindent}
  {1\@bls plus 0.25\@bls}{0.5em}{\normalsize\bf}}
\def\subparagraph{\@startsection{subparagraph}{4}{\parindent}
  {1\@bls plus 0.25\@bls}{0.5em}{\normalsize\bf}}

\def\@sect#1#2#3#4#5#6[#7]#8{\ifnum #2>\c@secnumdepth
  \def\@svsec{}\else
  \refstepcounter{#1}\edef\@svsec{\csname the#1\endcsname.\hskip0.5em}\fi
  \@tempskipa #5\relax
  \ifdim \@tempskipa>\z@
    \begingroup
      #6\relax
      \@hangfrom{\hskip #3\relax\@svsec}{\interlinepenalty \@M #8\par}%
    \endgroup
    \csname #1mark\endcsname{#7}\addcontentsline
      {toc}{#1}{\ifnum #2>\c@secnumdepth \else
        \protect\numberline{\csname the#1\endcsname}\fi #7}%
  \else
    \def\@svsechd{#6\hskip #3\@svsec #8\csname #1mark\endcsname
      {#7}\addcontentsline{toc}{#1}{\ifnum #2>\c@secnumdepth \else
        \protect\numberline{\csname the#1\endcsname}\fi #7}}%
  \fi \@xsect{#5}}

\long\def\@makefigurecaption#1#2{\vskip 10mm #1. #2\par}

\long\def\@maketablecaption#1#2{\hbox to \hsize{\parbox[t]{\hsize}
  {#1 \\ #2}}\vskip 0.3ex}

\def\fnum@figure{Figure \thefigure}
\def\figure{\let\@makecaption\@makefigurecaption \@float{figure}}
\@namedef{figure*}{\let\@makecaption\@makefigurecaption \@dblfloat{figure}}

\def\table{\let\@makecaption\@maketablecaption \@float{table}}
\@namedef{table*}{\let\@makecaption\@maketablecaption \@dblfloat{table}}

\textfloatsep=\floatsep           % Space between main text and floats
\intextsep=\floatsep              % Space between in-text figures and
\long\def\@makefntext#1{\parindent 1em\noindent\hbox{${}^{\@thefnmark}$}#1}

\def\maketitle{\begingroup        % Initialize generation of front-matter
    \def\thefootnote{\fnsymbol{footnote}}%
    \newpage \global\@topnum\z@
    \@maketitle \@thanks
  \endgroup
  \let\maketitle\relax \let\@maketitle\relax
  \gdef\@thanks{}\let\thanks\relax
  \gdef\@address{}\gdef\@author{}\gdef\@title{}\let\address\relax}

\def\justify@on{\let\\=\@normalcr
  \leftskip\z@ \@rightskip\z@ \rightskip\@rightskip}

\newbox\fm@box                    % Box to capture front-matter in

\def\@maketitle{%                 % Actual formatting of \maketitle
  \global\setbox\fm@box=\vbox\bgroup
    \vskip 8mm                    % 930715: 8mm white space above title
    \raggedright                  % Front-matter text is ragged right
    \hyphenpenalty\@M             % and is not hyphenated.
    {\Large \@title \par}         % Title set in larger font.
    \vskip\@bls                   % One line of vertical space after title.
    {\normalsize                  % each author set in the normal
     \@author \par}               % typeface size
    \vskip\@bls                   % One line of vertical space after author(s).
    \@address                     % all addresses
  \egroup
  \twocolumn[%                    % Front-matter text is over 2 columns.
    \unvbox\fm@box                % Unwrap contents of front-matter box
    \vskip\@bls                   % add 1 line of vertical space,
    \unvbox\abstract@box          % unwrap contents of abstract boxes,
    \vskip 2pc]}                  % and add 2pc of vertical space

\newcounter{address}
\def\theaddress{\alph{address}}
\def\@makeadmark#1{\hbox{$^{\rm #1}$}}

\def\address#1{\addressmark\begingroup
  \xdef\@tempa{\theaddress}\let\\=\relax
  \def\protect{\noexpand\protect\noexpand}\xdef\@address{\@address
  \protect\addresstext{\@tempa}{#1}}\endgroup}
\def\@address{}

\def\addressmark{\stepcounter{address}%
  \xdef\@tempa{\theaddress}\@makeadmark{\@tempa}}

\def\addresstext#1#2{\leavevmode \begingroup
  \raggedright \hyphenpenalty\@M \@makeadmark{#1}#2\par \endgroup
  \vskip\@bls}

\newbox\abstract@box              % Box to capture abstract in

\def\abstract{%
  \global\setbox\abstract@box=\vbox\bgroup
  \small\rm
  \ignorespaces}
\def\endabstract{\par \egroup}

\def\thebibliography#1{\section*{REFERENCES}\list{\arabic{enumi}.}
  {\settowidth\labelwidth{#1.}\leftmargin=1.67em
   \labelsep\leftmargin \advance\labelsep-\labelwidth
   \itemsep\z@ \parsep\z@
   \usecounter{enumi}}\def\makelabel##1{\rlap{##1}\hss}%
   \def\newblock{\hskip 0.11em plus 0.33em minus -0.07em}
   \sloppy \clubpenalty=4000 \widowpenalty=4000 \sfcode`\.=1000\relax}

\newcount\@tempcntc
\def\@citex[#1]#2{\if@filesw\immediate\write\@auxout{\string\citation{#2}}\fi
  \@tempcnta\z@\@tempcntb\m@ne\def\@citea{}\@cite{\@for\@citeb:=#2\do
    {\@ifundefined
       {b@\@citeb}{\@citeo\@tempcntb\m@ne\@citea
        \def\@citea{,\penalty\@m\ }{\bf ?}\@warning
       {Citation `\@citeb' on page \thepage \space undefined}}%
    {\setbox\z@\hbox{\global\@tempcntc0\csname b@\@citeb\endcsname\relax}%
     \ifnum\@tempcntc=\z@ \@citeo\@tempcntb\m@ne
       \@citea\def\@citea{,\penalty\@m}
       \hbox{\csname b@\@citeb\endcsname}%
     \else
      \advance\@tempcntb\@ne
      \ifnum\@tempcntb=\@tempcntc
      \else\advance\@tempcntb\m@ne\@citeo
      \@tempcnta\@tempcntc\@tempcntb\@tempcntc\fi\fi}}\@citeo}{#1}}

\def\@citeo{\ifnum\@tempcnta>\@tempcntb\else\@citea
  \def\@citea{,\penalty\@m}%
  \ifnum\@tempcnta=\@tempcntb\the\@tempcnta\else
   {\advance\@tempcnta\@ne\ifnum\@tempcnta=\@tempcntb \else \def\@citea{--}\fi
    \advance\@tempcnta\m@ne\the\@tempcnta\@citea\the\@tempcntb}\fi\fi}

\def\ps@crcplain{\let\@mkboth\@gobbletwo
     \def\@oddhead{\reset@font{\sl\rightmark}\hfil \rm\thepage}%
     \def\@evenhead{\reset@font\rm \thepage\hfil\sl\leftmark}%
     \let\@oddfoot\@empty
     \let\@evenfoot\@oddfoot}

\ps@crcplain                    % modified 'plain' page style

\documentstyle[twoside,fleqn,nuclphys,epsf]{article}
\input{psfig}

\def\aachen{{\it Proceedings of the Large Hadron
Collider Workshop}, edited by G. Jarlskog and D. Rein, Aachen (1990),
CERN 90-10, ECFA 90-133}
\def\prdj#1{{\it Phys. Rev.} {\bf D{#1}}}
\def\npbj#1{{\it Nucl. Phys.} {\bf B{#1}}}
\def\prlj#1{{\it Phys. Rev. Lett.} {\bf {#1}}}
\def\plbj#1{{\it Phys. Lett.} {\bf B{#1}}}
\def\zpcj#1{{\it Z. Phys.} {\bf C{#1}}}

\def\eg{{\it e.g.}}

\def\epem{e^+e^-}
\def\lsim{{\buildrel < \over\sim}}
\def\gsim{{\buildrel > \over\sim}}
\def\slash#1{#1\hskip-6pt/\hskip2pt}

\def\etmiss{\slash E_T}
\def\mmissl{M_{miss-\ell}}
\def\ie{{\it i.e.}}
\def\ebtag{e_{b-tag}}
\def\emistag{e_{mis-tag}}
\def\gam{\gamma}
\def\ell{l}
\def\nsd{N_{SD}}
\def\anti{\overline}
\def\pbi{~{\rm pb}^{-1}}
\def\fbi{~{\rm fb}^{-1}}

\def\gev{\,{\rm GeV}}
\def\tev{\,{\rm TeV}}

\def\wt{\widetilde}

\def\rta{\rightarrow}

\def\gl{\wt g}
\def\mgl{m_{\gl}}
\def\stop{\wt t}

\def\mstop{m_{\stop}}

\def\hsm{\phi^0}
\def\mhsm{m_{\hsm}}
\def\hl{h^0}
\def\hh{H^0}
\def\ha{A^0}
\def\hp{H^+}
\def\hm{H^-}
\def\hpm{H^{\pm}}
\def\mhl{m_{\hl}}
\def\mhh{m_{\hh}}
\def\mha{m_{\ha}}

\def\mhpm{m_{\hpm}}
\def\tanb{\tan\beta}
\def\cotb{\cot\beta}
\def\mt{m_t}
\def\mb{m_b}
\def\mz{m_Z}
\def\mw{m_W}
\def\mgut{M_U}

\def\wp{W^+}

\def\cnone{\wt\chi^0_1}
\def\cntwo{\wt\chi^0_2}
\def\snu{\wt\nu}

\def\msnu{m_{\snu}}
\def\mcnone{m_{\cnone}}
\def\mcntwo{m_{\cntwo}}
\def\h{h}
\def\mh{m_{\h}}
\def\cpone{\wt \chi^+_1}

\def\cpmone{\wt \chi^{\pm}_1}

\def\mcpmone{m_{\cpmone}}

\def\AmS{{\protect\the\textfont2
  A\kern-.1667em\lower.5ex\hbox{M}\kern-.125emS}}

\title{Searching for the Higgs Boson(s)
        \thanks{
   To appear in the Proceedings of the Zeuthen Workshop on Elementary
   Particle Theory --- {\it LEP 200 and Beyond}, Teupitz/Brandenburg,
   Germany, 10-15 April (1994), eds. T. Riemann and J. Blumlein.}
       }

\author{J.F. Gunion\address{Davis Institute for High Energy Physics,
        University of California at Davis, %\\
        Davis, CA 95616, U.S.A. }%
       }
\begin{document}

\begin{abstract}
The ability of LEP-200, the Tevatron, the Di-Tevatron, the LHC,
and a next linear $\epem$ collider (NLC) to probe the Higgs
sectors of the minimal Standard Model (SM) and
the Minimal Supersymmetric Model (MSSM) is reviewed.
Emphasis is placed on newly developed detection modes and on
predictions for supersymmetric particle Higgs decay channels
when the MSSM is constrained using boundary conditions
motivated by supergravity/superstring models.
\end{abstract}

\maketitle

\section{INTRODUCTION}

In assessing our ability to detect Higgs bosons at present and
future colliders, two prototype models are generally employed,
namely the Standard Model and the Minimal Supersymmetric Model.
While these certainly do not exhaust the theoretical or phenomenological
possibilities, they are probably the most attractive and compelling
models. The SM is the very simplest model in which electroweak
symmetry breaking (EWSB) can arise perturbatively; the only addition
to the particle menagerie is a single neutral Higgs boson ($\hsm$).
However, the SM has many free parameters and aesthetic problems
associated with naturalness and hierarchy. The MSSM provides
a dramatic cure for the latter, while simultaneously predicting
gauge-coupling unification consistent with experiment \cite{unif},
thereby linking physics at the unification
scale, $M_U$, to physics below a $\tev$. In order for this unification
to be successful, the specification of the MSSM that there be exactly two
Higgs doublets is crucial.
(A supersymmetric extension of the Standard Model
must have at least two Higgs fields
to give mass to both up and down quarks, and more than two doublets
spoils the unification.) Although it is possible for Higgs singlet
fields to be present without affecting the gauge-coupling unification,
the MSSM choice of no singlet Higgs is both minimal and attractive.
The resulting physical Higgs spectrum consists of two CP-even
scalars ($\hl$ and $\hh$ with $\mhl<\mhh$), a CP-odd scalar ($\ha$)
and a charged Higgs pair ($\hpm$).

In order for the SM to be perturbatively consistent,
$\mhsm<600\div 800\gev$ is required \cite{hhg}.  We will
focus on this mass region. In the MSSM, there are
many constraints on the masses of the physical Higgs bosons.
In both models, the branching ratios of the Higgs boson(s)
are completely determined once the masses of possible decay products
are fixed.  Reviews of these models can be found in Refs.~\cite{hhg}
and \cite{gunperspective}. This talk will presume a working knowledge
of all but recent results.

Clearly, to fully test the models will require verification of all
the predicted properties of the Higgs sector.
However, even detection of the Higgs bosons is by no means
straightforward, especially at hadron colliders.
In this talk, we focus on the following areas: a) What's `new' in
the SM $\hsm$ search; b) Progress in detecting the MSSM Higgs bosons;
c) The role of a 4 TeV `Di-Tevatron' in Higgs searches;
and d) the RGE/Superstring unification
context for the MSSM, focusing on special decay complications
that could emerge. These topics overlap somewhat, and so
it will be easiest to simply discuss the SM and the MSSM in turn.
The discussion will presume that the new CDF result of
$\mt\sim 170\gev$ \cite{CDF} is at least approximately correct.

\section{THE STANDARD MODEL $\bf \hsm$}

We discuss several related topics: detecting the $\hsm$ in
its $b\anti b$ decay mode; the influence of large $\mt$
on discovery modes at the LHC; and the possible role of a $4\tev$
Tevatron upgrade.

\subsection{Higgs detection in the $\bf b\anti b$ decay mode}

Two production/detection channels have been examined.
The first, primarily of relevance for LHC energies, is
$t\anti t\hsm\rta \ell b\anti b b\anti b X$ \cite{dgvttbb},
where we imagine tagging 3 or 4 of the $b$-quarks in association
with the leptonic trigger.
Irreducible backgrounds arise from $t\anti t b\anti b$ and $t\anti t Z$
(with $Z\rta b\anti b$)
production and reducible backgrounds from $t\anti t (g)$ production
(where the two $b$-quarks from the $t$'s are tagged and 1 (or 2) light
quark or gluon jet(s) is mis-tagged, \ie\ mis-identified, as a $b$-jet).
Various cuts and selection criteria for the trigger lepton and
$b$-jets are imposed.
The $b\anti b$ mass distribution (including all combinatoric effects)
is then constructed and the visibility of the $\hsm$ mass peak evaluated.
The statistical significance of the mass peak depends critically upon
the probability of tagging a given $b$-jet, $\ebtag$,
and the probability that a light quark or gluon jet is mis-tagged
as a $b$-jet, $\emistag$. Taking $\ebtag\sim 30\%$
(for $p_T^b\geq 20\div 30\gev$
and $|\eta^b|<2.5$) and $\emistag\sim 1\%$ (over the same kinematic range),
and requiring 3 $b$-tags,
Ref.~\cite{dgvttbb} finds that for $\mt\sim 170\div 180\gev$ a statistical
significance of $\nsd=5$ can be achieved after 0.2, 0.6, 1.8
LHC $100\fbi$ years for $\mhsm=80$, 100, and $120\gev$, respectively.
Tagging 4 $b$'s yields $\nsd=5$ after 0.7, 1.2 and 2.9 years at
these respective masses. Although, more integrated luminosity is
required, the 4 $b$-tag Higgs peaks are extremely clean for
the assumed $\emistag$ probability.

The second channel is $W\hsm\rta \ell b\anti b X$,
with irreducible background from $Wb\anti b$, $WZ$ and $t\anti b$ production,
and reducible backgrounds from $Wjj$ and $t\anti t$ production.
This has been examined
for Tevatron, Di-Tevatron and LHC energies in Ref.~\cite{stangeetalwhsm}
and for Di-Tevatron energies in Ref.~\cite{gunhan}. Double $b$-tagging is
employed, using the above-stated efficiencies. Observability of a $b\anti b$
mass peak at the $\nsd=5$ level is possible for $\mhsm\lsim70\gev$
at the Tevatron (with $L=10\fbi$). This means that
the Tevatron cannot compete with
LEP-II, which will reach $\mhsm\sim 80\gev$ even for $\sqrt s=176\gev$
(assuming $L=500\pbi$). At the LHC, the $W\hsm$ mode can probe up to
$\mhsm\lsim 120\gev$ (for $L=100\fbi$).
(Di-Tevatron results will be summarized shortly.)
Note that the mass reaches of the  $t\anti t\hsm$ and $W\hsm$
modes are quite similar at the LHC.  This complementarity would
prove very valuable in confirming these rather difficult signals.

At the LHC, the main issue will be whether
or not the LHC detectors can achieve the required $\ebtag$
and $\emistag$ levels in a high instantaneous luminosity, multiple
interaction environment. This is under study by the ATLAS and CMS
groups.  Early guesses are \cite{denegripc} that $\ebtag\sim 30\%$
can be achieved, but that $\emistag$ as low as $1\%$ will prove
difficult unless a design for the vertex detector allowing placement
close to the interaction point can be found.  This is a critical issue
for the LHC detectors. The $t\anti t \hsm$ and $W\hsm$ associated production
modes both require good $b$-tagging efficiency and purity and are the only
modes allowing access to the $b\anti b$ mode that dominates $\hsm$
decays for $\mhsm\lsim 140\gev$. If $b$-tagging is only possible
at ${\cal L}\sim 10^{33} cm^{-2}sec^{-1}$, corresponding to $L=10\fbi$
per year, then the $t\anti t\hsm$ and $W\hsm$ modes will only
probe up to $\mhsm\sim 75$ ($95$) GeV after 1 (3) years.

\subsection{Impact of large $\bf \mt$ at the LHC}

The fact that the top quark is heavy impacts the search for the $\hsm$
at the LHC by yielding a large cross section for $t\anti t\hsm$
associated production along with relatively
small $t\anti t$-related irreducible
backgrounds. It is this which makes the $\hsm\rta b\anti b$
signal in the $t\anti t\hsm$ process competitive with that from the $W\hsm$
reaction.  Another important discovery mode is
the $t\anti t \hsm \rta \ell\gam\gam X$
channel \cite{lgamgam}, with irreducible background from
$t\anti t \gam\gam$ production \cite{ttbargamgam}.
The $\ell\gam\gam$ channel becomes much cleaner \cite{summers} at high
$\mt$. Increasing $\mt$ increases the signal while decreasing the
$t\anti t\gam\gam$ background.
At $\mt=180\gev$, a $\mhsm=100\gev$ Higgs yields about 32 events
at a $\sqrt s=16 \tev$ LHC compared to $\sim 2$ events
from $t\anti t \gam\gam$ (for integrated luminosity of $L=100\fbi$
and 3\% $\gam\gam$ mass resolution). The nominal statistical significance,
(defined by $\nsd\equiv S/\sqrt B$, where $S$ and $B$ are the
signal and background rates, respectively)
of $\nsd\sim 21$ will, of course, be reduced by imperfect jet/photon
discrimination and so forth, but detection of the $\hsm$
in this channel should be straightforward for
$60\div 70\lsim\mhsm\lsim 130\div 140 \gev$.

\subsection{A Di-Tevatron}

With the cancellation of the SSC, a critical issue for the U.S. high
energy physics community has become whether or not to push a
number of possible upgrades for the Tevatron or to participate in
a major way at the LHC. In this context, the possibility of
a Di-Tevatron with $\sqrt s \sim 4\tev$ and yearly luminosity
of $L=10\div 30\fbi$ has been discussed.  Several groups have examined
$\hsm$ detectability at such a machine.
The first mode of interest is $W\hsm\rta \ell b\anti b$ associated production
\cite{stangeetalwhsm,gunhan} ---
$t\anti t\hsm$ production has too low a rate at this low energy
to be of use. Although there are a few differences in details
of the treatments (such as inclusion
of the relatively small $t\anti b$ backgrounds
in Ref.~\cite{stangeetalwhsm},
and inclusion in Ref.~\cite{gunhan} of peak broadening due
to semi-leptonic decays) the basic conclusions are very similar.
The best signals are achieved by requiring 2 $b$-tags; values
of $\ebtag=30\%$ and $\emistag=1\%$ (as above) are assumed.
At $\sqrt s=3.5\tev$ and $L=30\fbi$,  Ref.~\cite{stangeetalwhsm}
concludes that $\nsd\gsim 5$ can be achieved
for $\mhsm\lsim 100\gev$ assuming conservative resolutions
and cuts, while Ref.~\cite{gunhan} finds that
at $\sqrt s=4\tev$ and $L=30\fbi$, $\nsd\gsim 5$ can be achieved
for $\mhsm\lsim 110\div 125\gev$, the lower (upper) limit applying
for conservative (optimistic) resolutions and cuts.

%ffffffffffffffffffffffffffffffffffffffffffffffffffffff
\vspace{.5cm}
\begin{figure}[htbp]
\begin{center}
%\mbox{
%\epsfysize=7.5cm
%\epsffile{zeuthen_4l4tev.ps}
%}
%\mbox{
\leavevmode
{\psfig{figure=zeuthen_4l4tev.ps,width=3in,clip=}}
%     }
\end{center}
\caption{
The $gg\rta \hsm\rta ZZ^{(*)}\rta 4l$ ($l=e,\mu$) signal and
$q\anti q \rta ZZ \rta 4l$ background as a function of the four-lepton mass,
$M_{4l}$, in 5 GeV bins.  Higgs boson signals for $\mhsm=130$, 150, 170, 200,
230, 270, 300 and $400\gev$ are illustrated.}
\label{4l4tevfig}
\end{figure}
%------------------------------------------------------------

A variety of modes that might allow observation of the $\hsm$
at the Di-Tevatron for $\mhsm\gsim 120\gev$ were explored
in Ref.~\cite{gunhan}.  The cleanest channel is
$\hsm\rta ZZ^{(*)}\rta 4\ell$.  Adequate event number is critically
dependent upon being able to employ relatively soft cuts for
the $e$'s and $\mu$'s ($p_T^{\ell}\geq 10\gev$ in $|\eta^{\ell}|<2.5$).
The resulting $4l$ mass spectra for a variety of Higgs
masses are shown in Fig.~\ref{4l4tevfig}, assuming $L=30\fbi$.
Consider first the region $\mhsm\lsim 2\mz$. We see
that an observable signal is present only
in the vicinity of $\mhsm\sim 150\gev$, for which mass one obtains
$\sim 13$ (background free) events. On either side of $\mhsm\sim 150\gev$,
the event rate declines to a problematical level. For instance,
for $\mhsm\sim 130$ or 170 GeV, only some 5 or 4 events are predicted,
probably too few to claim discovery.

In the $\mhsm\gsim 2\mz$ region, there is background
from continuum $ZZ\rta 4\ell$ production, as shown in Fig.~\ref{4l4tevfig}.
Nonetheless,
a $\nsd=5$ signal is predicted for $2\mz\lsim \mhsm\lsim 220\gev$.
Because of the cleanliness of the channel, a signal at
the $\nsd\sim 3$ level might be deemed acceptable, in which case one
can reach to $\mhsm\lsim 270\div 300\gev$. To reach $\nsd\sim 5$
at such masses requires $L\sim 60\div 100\fbi$. If
the detector-dependent backgrounds in the $Z+\etmiss$ channel
can be kept small, then  for $\mhsm>2\mz$
the $ZZ\rta 2\ell 2\nu$ mode would yield
a confirming signal at a similar $\nsd$ level to that found
in the $4\ell$ mode. Signals at the $\nsd=5$ level
in the $\hsm\rta WW\rta l\nu j j$ channel are also predicted
for $150\lsim \mhsm\lsim 225\gev$ for $L=30\fbi$.

Thus, a Di-Tevatron upgrade would allow some access to the SM Higgs
sector, but $\mhsm$ could fall outside the mass ranges for
which discovery would be possible.  Model builders with
a bias towards a light $\hsm$ would be especially concerned
about the $110\div 150\gev$ discovery gap. Unfortunately,
the Di-Tevatron is not able to extend much (if at all)
beyond the $\mhsm\lsim 105\gev$ discovery limit
of LEP-II when operated at $\sqrt s=200\gev$. Of course,
if LEP-II only reaches $\sqrt s=176\gev$, for which only $\mhsm\lsim 80\gev$
can be probed, the Di-Tevatron would provide a significant addition
to the range of light $\hsm$ masses that could be probed.

\section{MINIMAL SUSY HIGGS BOSONS}

The assessment of prospects for detection of the MSSM Higgs bosons is far more
complex than for the SM $\hsm$. At tree-level, we recall that
all properties of the Higgs sector are determined by specification
of just two parameters: normally one employs $\mha$ and $\tanb$
(the standard ratio of vacuum expectation values). At tree-level one finds
$\mhl\leq \mz$, $\mhh\geq\mz$, and $\mhpm\geq \mw$, with $\mhl$ reaching its
maximum at large $\mha$ and large $\tanb$ while the minimum $\mhh$
and $\mhpm$ values are approached at small $\mha$ \cite{hhg}.  However,
the tree-level Higgs masses receive radiative
corrections \cite{habperspective}. For $\mt$ as large as $\sim 170\gev$,
the $\hl$ mass can be greatly enhanced if the mass of the stop quark,
$\mstop$, is also large --- the shift in mass squared is roughly given by
\begin{equation}
\Delta \mhl^2\sim {3 G_F\over \pi^2\sqrt 2\sin^2\beta}\mt^4
\ln {\mstop^2\over \mt^2}\,
\label{mhlmaxeq}
\end{equation}
where we have neglected squark mixing. The result (at large $\tanb$)
is $\mhl^{max}\sim 115$, 125, 138 GeV for $\mt=150$, 170, 190 GeV,
assuming $\mstop=1\tev$.
The minimum value of $\mhh$ is increased above $\mz$ by a like amount.
Finally, it will also be important to keep in mind the approximate
degeneracy, $\mha\sim\mhh\sim\mhpm$, that holds once $\mha\gsim 2\mz$.

Current experimental constraints on the MSSM Higgs bosons from LEP are
roughly $\mhl\gsim 40\gev$ and $\mha\gsim 20\gev$ \cite{pdg}.
The LEP bound of $\mhpm\gsim 44\gev$ is still below the minimum
expected for $\mhpm$ in the MSSM.
The CDF top-quark observation \cite{CDF} provides
only very weak constraints on $\mhpm$. What constraints there are
derive from the fact that the large observed event rate in the standard
$t\rta \wp b$ decay channel makes it unlikely that the
decay $t\rta \hp b$ could have a substantial branching ratio.
However, for $\mt$ as large as 170 GeV, $BR(t\rta \hp b)\gsim 0.2$
only for rather small or very large $\tanb$ values.
The giant dip in the branching ratio
for $\tanb\sim 6\div 7$ as the $\hp\rta t\anti b$ coupling switches
from $\mt\cotb$ to $\mb\tanb$ dominance keeps the predicted
branching ratio small in the most interesting region,
$2\lsim\tanb\lsim 20$ \cite{hhg}.

%ffffffffffffffffffffffffffffffffffffffffffffffffffffff
\vspace{.5cm}
\begin{figure}[htbp]
\begin{center}
%\mbox{
%\epsfysize=7.5cm
%\epsffile{zeuthen_lepii.ps}
%}
%\mbox{
\leavevmode
{\psfig{figure=zeuthen_lepii.ps,height=3.5in,width=3in,clip=}}
%     }
\end{center}
\caption{Contours for $Z\hl$ detection at LEP-II, assuming $\mha=1\tev$.
To the right of the contours, fewer than 100 $Z\hl$
events are predicted for $L=500\pbi$.}
\label{lepiifig}
\end{figure}
%------------------------------------------------------------

\subsection{LEP-II and NLC Prospects}

The impact of large $\mt$ on observation of $Z\hl$ associated
production at LEP-II is especially dramatic.  Assuming that $\mha\gsim 2\mz$
(highly likely in most MSSM unification scenarios, as we shall later
outline), we can compute $\mhl$ as a function of
$\mt$, $\tanb$ and $\mstop$.
We then determine the maximum value of $\mstop$ as a function of $\mt$
that yields $\geq 100$ $Z\hl$ events for a given $\tanb$ value and
given LEP-II energy, assuming an integrated luminosity of $L=500\pbi$.
Results for various fixed $(\sqrt s,\tanb)$ values,
from Refs.~\cite{gunperspective,gunhawaii}, appear in
Fig.~\ref{lepiifig}. (See also Ref.~\cite{janotlep}.)

For example, consider $\mt=170\gev$.
For $\sqrt s=175\gev$, 100 $Z\hl$ events are predicted
for $\mstop~\lsim 500$ ($150$) GeV if $\tanb\sim 2$
($\tanb\gsim 20$), while at $\sqrt s=200\gev$, these respective
$\mstop$ values are 1800 (400) GeV. In a large class
of superstring and supergravity models the $\stop$ is relatively
light, $\mstop~\lsim 300\div 500\gev$, while $\tanb$
can take on values ranging from 2 to 20.  As a result, $\hl$
detection in such models is {\it far} more likely at LEP-200 than at LEP-176
for this CDF-like value of $\mt$.
We also note from Fig.~\ref{lepiifig} that the maximum $\mstop$
that would allow $Z\hl$ detection at a given $(\sqrt s,\tanb)$
can decrease very dramatically from quite large to quite small values
as $\mt$ increases above a certain critical value.

In considering the $Z\hl$ mode, it is important to keep in mind
that the $\hl$ may decay to mostly invisible channels
such as $\cnone\cnone$ \cite{habinv} or $\snu\snu$.
In fact, we shall later see that unified string/supergravity
models predict that such decays {\it will} dominate
if the gluino mass (which sets the scale of $\mcnone$ and $\msnu$)
is near the lower limited allowed by theoretical and current experimental
constraints. At an $\epem$ collider, invisible decays do not
present a particular problem since the visible $Z$ decays
along with the beam constraints allow reconstruction of the missing mass
spectrum.

Of course, at an NLC with $\sqrt s\gsim 300\gev$ detection of $Z\hl$
is guaranteed \cite{gunperspective,gunhawaii,brignole,zerwas,janothawaii};
even if the SUSY model
is extended to include extra singlets and all parameters
are pushed to their perturbative limits,
one finds $\mhl\lsim 150\gev$ \cite{kanehl}.
The important issue at the NLC is detection of the $\ha$, $\hh$
and $\hpm$.  Kinematics are an especially limiting factor
on the maximum masses that can be probed due to the fact that
the only substantial production processes at large $\mha\sim\mhh\sim\mhpm$
are $\epem\rta \ha\hh$ and $\epem\rta \hp\hm$.
One finds that masses up to and slightly beyond $0.4\sqrt s$
will allow reasonable rates for these pair production processes,
assuming an appropriate integrated
luminosity \cite{gunperspective,gunhawaii,brignole,zerwas,janothawaii}.
Thus, for example at $\sqrt s=500\gev$ and $L=10\fbi$,
the $\ha$, $\hh$ and $\hpm$
can all be detected for masses $\lsim 210\div 220\gev$.
Unfortunately, we will find that this is rather limiting.  In many
unified model contexts the lower bound on $\mha$
for the allowed parameter space is $\sim200\gev$, and the bulk
of parameter space corresponds to substantially higher $\mha$ values.

It has been suggested \cite{gunhabgamgam,gunhawaii} that operating
the NLC in the back-scattered laser beam $\gam\gam\rta$Higgs collision mode
could increase the $\mha$ and/or $\mhh$ mass reach. The extent to which this
works is strongly dependent upon $\tanb$ and upon the decay modes
of the $\ha$ and $\hh$. If SUSY decay modes are forbidden,
and assuming an integrated $\gam\gam$ luminosity of $20\fbi$
and the availability of polarized laser and $\epem$ beams,
it is estimated that $\mha$ and $\mhh$ masses as high as $0.8\sqrt s$
could be probed if $\tanb$ has a moderate value ($\tanb\lsim 4$).
The channels $\hh\rta \hl\hl$
and $\ha\rta Z\hl$ for $\mhh,\mha\lsim 2\mt$, and $\hh,\ha\rta t\anti t$
for $\mhh,\mha\gsim 2\mt$, will yield viable signals.
For large $\tanb$, the $b$-quark loop contribution to the $\gam\gam\rta\hh,\ha$
couplings becomes enhanced and the consequent cancellations suppress
the event rate. Mass reach is unlikely to exceed that achievable
via normal $\epem$ collisions.  If SUSY decays of the $\ha$
and $\hh$ are important (as is the case in many unified
supergravity and superstring models) detection in the $\gam\gam$
collision mode may prove even more difficult \cite{gunkelohn}.

\subsection{LHC Prospects}

We first review the early work leading to an approximate `no-lose'
theorem for LEP-II in combination with the LHC/SSC supercolliders.
We then discuss the impact on the `no-lose' theorem due to:
a) more recently developed detection modes; and b)
the complications that arise if SUSY decays of the MSSM Higgs
bosons are significant.

\subsubsection{Towards a no-lose theorem}

Early studies of the prospects for detecting
the MSSM Higgs bosons at the LHC (and the now-defunct SSC)
\cite{hhzz}-\cite{baeretal} focused on attempting to establish
a `no-lose' theorem, according to which at least one of the MSSM
Higgs bosons would be seen either using the $Z\hl$ mode at
LEP-II (with $\sqrt s=200\gev$, $L=500\pbi$
assumed) or at the LHC (SSC) with $L=100\fbi$ ($30\fbi$).
We focus on the LHC, and presume for now that the superpartner
particles are all too heavy to appear in the MSSM Higgs decays.
Supersymmetric decays could deplete the signals in the channels
discussed below; we will return to the effects of such decays.

%ffffffffffffffffffffffffffffffffffffffffffffffffffffff
\vspace{.5cm}
\begin{figure}[htbp]
\begin{center}
%\mbox{
%\epsfysize=7.5cm
%\epsffile{zeuthen_hole_smallsize.ps}
%}
%\mbox{
\leavevmode
{\psfig{figure=zeuthen_hole_smallsize.ps,width=3in,clip=}}
%     }
\end{center}
\caption{Discovery contours for LEP-200 ($L=500\pbi$) and LHC
($L=100\fbi$) for $\mt=150\gev$ and $\mstop=1\tev$.}
\label{contoursfig}
\end{figure}
%------------------------------------------------------------

As reviewed in, for example, Ref.~\cite{gunperspective},
the combined LHC discovery modes of
(i) $t\anti t\rta W^{\pm}H^{\mp} b\anti b
\rta \ell\tau b\anti b X$
\cite{barnettetal} (ii) $gg\rta\hl\rta \gam\gam$ \cite{gkw}
and $W\hl\rta \ell \gam\gam X$ plus
$gg\rta t\anti t \hl\rta \ell \gam\gam X$
\cite{whlgamgam,lgamgam} (iii) $gg\rta \hh\rta ZZ\rta 4\ell$
and (iv) $gg\rta \hl\rta ZZ^*\rta 4\ell$ \cite{gkw}
(viable for $\mhl\gsim 130\gev$) come close to
allowing Higgs discovery at the LHC over those sections of
the standard $\mha-\tanb$ parameter space for which $Z\hl$
discovery at LEP-200 is not possible. Because of the radiative
corrections to $\mhl$ and $\mhh$, the relative utility
of LEP-200 vs. the LHC is a strong function of $\mt$ and $\mstop$.
The two machines are highly complementary for $\mt\sim 150\gev$
and $\mstop~\gsim 800\div 1000\gev$.

This is illustrated in Fig.~\ref{contoursfig}. There we plot
the discovery contours for the LHC processes (i) and (ii) and
for $Z\hl$ detection at LEP-200, labeled by letters  `g', `e' and `a',
respectively --- ignore the other contours for now.
The letters are placed on the inside of the region for which detection
in the associated mode is possible.
For additional useful figures, which also display additional
detection modes, see Ref.~\cite{gunperspective} or
Refs.~\cite{gunorr,bargersusy,kzsecond,baeretal}.
For instance, there is a well-known tear-drop shaped
region with $\mha\lsim 2\mt$ at low to moderate $\tanb$
for which $\hh\rta 4\ell$ detection will be possible.

However, the modes (i)-(iv) do not provide complete coverage if
$\mt\sim 150\gev$ and $\mstop\sim 800\div 1000\gev$;
a `hole' is left in $\mha-\tanb$ parameter
space for $\mha\sim 100\div 175\gev$ and $\tanb\gsim 8\div 10$
for which neither LEP-200 nor LHC modes (i)-(iv) would allow
detection of any of the MSSM Higgs bosons.
This hole is marked by the big $X$
in Fig.~\ref{contoursfig}.  It is defined by
the LEP-200 $Z\hl$ and the LHC $t\rta \hp b$ and $\hl\rta \gam\gam$
discovery contours. To fill the hole, Ref.~\cite{kzfirst} raised
the possibility of detecting
$gg\rta \ha b\anti b$ and $gg\rta\hh b\anti b$ in the $\tau^+\tau^-b\anti b$
final state at large $\tanb$. This mode has been semi-realistically evaluated
by all of the LHC detector collaborations with positive results
for $\mha\gsim 100\gev$ and $\tanb\gsim 10$ \cite{felcini}.
(We remind the reader that high $\tanb$ enhances the $b\anti b$
coupling to $\hh$ and $\ha$, while at the same time
guaranteeing a branching ratio for $\ha,\hh\rta \tau^+\tau^-$
of order 10\% {\it assuming that there are no SUSY decay modes
of the $\ha$ or $\hh$}.) Thus, at $\mt\sim 150\gev$
we are left with only a small gap in
Higgs detection coverage in the region
$\mha\sim 100\div 175\gev$ and $\tanb\lsim 10$.

However, this close approach to a `no-lose' theorem is
quite sensitive to the LEP-II center-of-mass energy,
as is the degree of complementarity between LEP-II and the LHC.
Decreasing the LEP-II energy
to $\sqrt s=176\gev$ would extend the hole at $\mha\sim100\div 175\gev$ to
significantly lower $\tanb$ values, well below the reach of
the $\tau^+\tau^-$ mode. Conversely, increasing the LEP-II energy
to about 230 GeV would allow $Z\hl$ detection for all of parameter space
assuming $\mstop~\lsim 1 \tev$.

The relative importance of LEP-II vs. the LHC
is sensitive to $\mt$ because as $\mt$ is increased
the maximum value of $\mhl$ (which is rapidly approached in the
$\mha\gsim 2\mz,\tanb\gsim 3$ corner of parameter space)
and the lower limit for $\mhh$ (approached at $\mha\lsim \mz$)
both increase. This has several crucial effects.  We describe
results for $\mt\sim 200\gev$ and $\mstop\sim 1\tev$.
First, $\mhl$ is increased sufficiently that the $Z\hl$ mode at LEP-200
remains viable at low $\mha$ only for $\tanb\lsim 3$ and not at all
for $\mha\gsim 250\gev$. Fortunately, detection of at least one
of the MSSM Higgs bosons at the LHC becomes possible for essentially
all of $\mha-\tanb$ parameter space. (A) The larger $\mhh$ value at low $\mha$
opens up the region for which $\hh\rta 4\ell$ will be possible
--- indeed $\hh\rta ZZ^*\rta 4\ell$ is viable for essentially
all $\mha\lsim 130\gev$, while the $\hh\rta ZZ\rta 4\ell$ mode
yields $\nsd\geq 4$ over an expanded region at higher $\mha$ and moderate
to low $\tanb$. (B) The increased $\mhl$ values allow observable
$\hl\rta ZZ^*\rta 4\ell$ rates for a large region described roughly
by $\mha\gsim 150\gev$ and $\tanb\gsim 1.5$.
(C) Increasing $\mt$ also increases the portion of parameter space
covered by the $t\anti t$ charged Higgs final state mode (i) to
roughly $\mha\lsim 150\gev$.  (D) The region of viability for $\hl$ detection
in the $\gam\gam X$ and $\ell\gam\gam X$ modes
remains about the same as shown in Fig.~\ref{contoursfig}. The result,
even before inclusion of the $b\anti b\tau^+\tau^-$ mode,
is that the combination of all the above modes leaves only a tiny wedge
of questionable coverage at $\mha\sim 140\gev$ and $\tanb\gsim 6\div 7$.

The CDF top quark mass value of $\sim 170\gev$ lies between
the above two cases. It is such that if $\mstop$ is also large then
both LEP-200 and the LHC will be needed to have a high probability
to detect an MSSM Higgs boson, without employing the new detection
modes to be discussed shortly.

Finally, we discuss the very
strong dependence of the relative utility of LEP-200 vs. the LHC
on $\mstop$. This is an important consideration in view of the fact
that many very attractive unification treatments of the MSSM
employ boundary conditions that give rise to low $\mstop$ values ---
\ie\ from a theoretical point of view
low $\mstop$ is at least as likely as large $\mstop$.
If $\mstop~\lsim 300\gev$ (a not atypical value for many models),
LEP-200 will play a dominant role;
at $\sqrt s=200\gev$  the $Z\hl$ or $\ha\hl$ processes
will be visible for all of $\mha-\tanb$ parameter space
for any $\mt\leq 200\gev$. Meanwhile, such a low value of $\mstop$
will imply deterioration of those LHC signals that require
large maximum $\mhl$ or large minimum $\mhh$ values; these include
the $\hh,\hl\rta ZZ^*\rta 4\ell$ signals in particular, with
the $\hl\rta\gam\gam$, $t\anti t\hl\rta \ell 3bX$ and $W\hl\rta \ell 2bX$
modes also displaying sensitivity.

\subsubsection{New LHC detection modes}

There are a number of processes in which MSSM Higgs detection in
$b$-quark decay modes would be possible if high efficiency
and purity for $b$-tagging
can be achieved in a high instantaneous luminosity
environment.  We have already discussed $t\anti t+$Higgs
and $W+$Higgs
production in the case of the $\hsm$.  The useful cases in the
MSSM are: $t\anti t\hl$, $t\anti t \hh$, $W\hl$ and $W\hh$.
(The $\ha$ analogues do not generally produce viable signals.
For $\tanb>1$, $t\anti t \ha$ is suppressed because of the
$\mt\cotb$ form of the coupling; $W\ha$ production via $W^*$
is suppressed by the absence of a tree-level $WW\ha$ coupling.)
At high $\tanb$, there are several additional channels that
have the potential to yield an observable Higgs signal.
These are the $gg$ fusion produced final states of
$b\anti b +\hh,\ha,\hl$ (with $\hh,\ha,\hl\rta b\anti b$)
and $b\anti t\hp+\anti b t \hm$ (with $\hp\rta t\anti b$
and $\hm\rta \anti t b$).
We review results for all the above modes below.

When the $\hl$ has mass below the range for which detection
in the $\gam\gam$ or $\ell\gam\gam$ final states is possible,
its branching ratio to $b\anti b$ is generally not much below unity.
Further, so long as $\mha\gsim \mz$ the $\hl$ has roughly SM-like
coupling to $t\anti t$, and its coupling and branching
ratio to $b\anti b$ can actually be somewhat
enhanced over SM-like results (at the same mass, $\mhsm=\mhl$).
Detection of $t\anti t \hl$ production
in the $\ell 3bX$ or $\ell 4bX$ final states
could prove possible. For $\ebtag=30\%$ and $\emistag=1\%$,
at $L=100\fbi$ the $t\anti t\hl\rta \ell 3b X$ process
produces $\nsd=4$ peaks in the $2b$ mass spectrum over a significant
portion of the $\mha-\tanb$ parameter space for the canonical
$\mt\sim150\gev$ and $\mstop\sim 1\tev$ mass choices \cite{dgvii}.
In particular, a statistically significant signal emerges
in much of the `hole' region found
at $\mt=150\gev$. The actual boundaries for the discovery region
appear (labelled {\it on the interior} by the letter `i')
in Fig.~\ref{contoursfig}. The precise boundary is quite
`soft' in many senses. The allowed discovery region
expands dramatically if only $\nsd=3$ is required, or (equivalently)
integrated luminosity is increased by a factor of 16/9; and it
contracts rapidly if the $\nsd$ requirement is increased.
This means that details of cuts, resolutions, inclusion of semi-leptonic
decays and tagging efficiencies and so forth are important.

Regarding tagging, the ability of the LHC detectors
to achieve the required $b$-tagging purity is far from certain
at high instantaneous luminosity, but the importance of reaching
the $\ebtag$ and $\emistag$ milestones needed for viability of the
$t\anti t \hl$ mode is great.
Indeed, we shall see that the 30\% and 1\% criteria
required will simultaneously open up the possibility
of obtaining useful signals in the other modes mentioned above,
to which we now turn.

The $W\hl\rta \ell b\anti b$ mode with double $b$-tagging and vetoing
against more than two jets (to kill $t\anti t$-related and other
backgrounds) was explored for the LHC in Ref.~\cite{stangeetalwhsm}.
They predict that for $L=100\fbi$ the $W\hl$ mode would yield $\nsd=5$
or better in a region extending roughly from the lower $\mha$
contour indicated by the letter `i'
for the $t\anti t \hl$ mode in Fig.~\ref{contoursfig}
to arbitrarily large $\mha$ values, \ie\ a more extensive region
than for the $t\anti t\hl$ mode. However, in comparing
these results we again caution that these contours are
somewhat soft and depend upon many details.
Most probably, the two techniques will provide very similar
$\hl$ detection capabilities.

At higher $\mt\sim 200\gev$, $\mhl$ becomes larger at a given
$\mha-\tanb$ point in parameter space.  This has two unfavorable
impacts on the $t\anti t \hl$ and $W\hl$ channels.
First, if $\mstop\sim 1\tev$
$\mhl$ can become large enough for all but small $\tanb$
that $\hl\rta WW^*$ decays cause a decrease in the
$\hl\rta b\anti b$ branching ratio. Second, the $W\hl$ and (at fixed $\mt$)
$t\anti t \hl$ cross sections decline with increasing $\mhl$.
The result is that if $\mt\sim 200\gev$ then $2b$-spectrum
mass peaks with $\nsd\geq 4$ arise in $t\anti t \hl$ associated production
only at low $\tanb$. Similarly, the results of Ref.~\cite{stangeetalwhsm} show
a greatly reduced region for which $\hl$ detection in the $W\hl$ mode
would be possible when $\mt\sim 190\gev$.  However, for 3 to 4 times
the standard $L=100\fbi$ integrated luminosity, the viability regions
for both the $t\anti t\hl$ and $W\hl$ modes expand even beyond
those found for $L=100\fbi$ at $\mt\sim 150\gev$.
Of course, if the stop squark
is light ($\mstop~\lsim 300\gev$), $\mhl$ would not receive
large radiative corrections, and both modes remain viable at $L=100\fbi$ even
for these larger $\mt$ values.

Detection of the $\hh$ in the $t\anti t \hh$ or $W\hh$
modes is also possible. In the case of $t\anti t \hh$, an $\nsd\geq 3.5$
(but not 4) signal is found at $\mt=150\gev,\mstop=1\tev$
for a region defined by $\mha\sim 50\div 100\gev$
and $\tanb\gsim 3$, \ie\ just to the left of the lower letter-`i'
contour of Fig.~\ref{contoursfig}. A similar result undoubtedly
applies for the $W\hh$ mode.

As mentioned at the beginning of this section,
at high $\tanb$ rates are enhanced for the processes
$gg\rta \ha b\anti b$ (for all $\mha$),
$gg\rta \hl b\anti b$ (for $\mha\lsim \mz$ where the $\hl$
and not the $\hh$ has enhanced couplings) and $gg\rta \hh b\anti b$
(for large $\mha$). A Higgs with enhanced $b\anti b$ coupling
is also more or less guaranteed to have $b\anti b$ branching
ratio of order 90\% (independent of mass)
in the absence of SUSY decay modes.
We have already noted that the large rates for $\hh b \anti b$
and $\ha b\anti b$ production lead to statistically significant signals
for the $\hh$ and $\ha$ in the $\tau^+\tau^-b\anti b$ channel
for large enough $\tanb$.
Correspondingly, when the above cross sections are enhanced
at large $\tanb$, the $b\anti b b\anti b$ final state
can be probed using
triple $b$-tagging (with $\ebtag=30\%$ and $\emistag=1\%$).
Statistically significant peaks are found in the $2b$ mass spectrum
\cite{dgvbbbb} --- 3 $b$-tags are required to reduce the
otherwise very large $b\anti b g$ background
to the level of the irreducible $b\anti b b\anti b$ background.
(As an aside, we note that {\it single} $b$-tagging will probably
increase the significance of the $\tau^+\tau^-$ signal for which
most backgrounds do not have an associated $b$-quark.)
The regions of $\mha-\tanb$ parameter space yielding $\nsd\geq 4$
(assuming $L=100\fbi$ of integrated luminosity)
for the $\hh$, $\hl$ and $\ha$ are indicated in Fig.~\ref{contoursfig}
by the boundaries marked with the letters `k', `l' and `m', respectively.
Note how large a section of parameter space would be covered
by these modes. At higher $\mt\sim 200\gev$, the general picture
for the $b\anti b b\anti b$ final state modes
does not change --- the only effect is to shift the transition
zone between $\hh$ viability and $\hl$ viability to somewhat
higher $\mha$.

However, there is an additional caveat concerning these modes
beyond the question of whether the required $\ebtag$ and $\emistag$
goals can be achieved.  Namely,
the statistical significance, $\nsd$, does not fully reflect
the difficulty of seeing the mass peak.  Typically, the $S/B$
ratio is of order 1\%, so that the shape of the background
over a region about twice the Higgs resolution-smeared width
must be quite well predicted in order to allow isolation
of the signal. In practice, the predicted background shape
would be adjusted to yield good agreement with the overall normalization
of the observed $b\anti b$ spectrum over an interval of order $30\div 40$ GeV,
and then a systematic deviation typical of a $15\div 20$ GeV width resonance
peak would be searched for. Obviously, systematic uncertainties
in the predicted background shape would have to be minimal ---
the detector would have to be well understood and the Monte Carlo very
finely tuned. Since the background exhibits a broad peak in
the vicinity of $M_{b\anti b}\sim 100\gev$, this would be more
of a problem for the $\hl$ or a light $\ha$. At higher masses,
the background falls very smoothly and monotonically
and would be less influenced by gradual mass-dependent
shifts in detector efficiency and the like. The ability
to detect a small peak of $\nsd=5$ nominal statistical
significance due to a heavy $\ha$ or $\hh$ would be relatively certain.
We are optimistic that the LHC detectors will be well enough
understood that simultaneous observation of
a statistically significant signal in the $\tau^+\tau^-$ and $b\anti b$
spectra will be possible if adequate $b$-tagging purity and efficiency
is available.

The mode $gg \rta t\anti b \hp + c.c.$ was studied in
Refs.~\cite{gunhp} and \cite{bargerhp}. Using 3-$b$-tagging
and reconstructing the $tb$ mass can produce highly significant
$\hp$ mass peaks provided $BR(\hp\rta t\anti  b)\sim 1$ and
the $\hp\rta t\anti b$ coupling
has the strength predicted at $\tanb=1$.
Unfortunately, the $\mt\cotb$ component of the
coupling strength declines very rapidly as $\tanb$ increases,
and it is only at quite large $\tanb$ that the $\mb\tanb$
component returns the net coupling to the original $\tanb=1$ level.
Even for 100\% $\hp\rta t\anti b$ branching ratio,
only for $\tanb\lsim 1.5$ and $\tanb\gsim 25\div 30$ does
one find an $\nsd\geq 5$ $tb$-spectrum mass peak.
While this seems a severe limitation, we note that some
unification treatments of the MSSM require
$\tanb\lsim 2$ or $\tanb\gsim 30\div 40$ if exact $\lambda_b=\lambda_\tau$
Yukawa unification is required at the unification scale and
if $\mt\lsim 180\gev$ \cite{pokorski}.
Of course, in these same unified models
it is also possible (but not necessary) that the branching ratio for
$\hp\rta t \anti b$ is substantially smaller than unity due to the presence
of SUSY decay modes.

\subsubsection{Influence of SUSY Decays}

As frequently noted in the previous section, all the above
results presume that the MSSM Higgs do not have significant decays
to SUSY-particle-pair modes.  This is almost certainly too
optimistic.  As hinted at several times, many models \cite{gunpois,bgkp},
including the unified models to
be discussed later, predict a host of relatively light supersymmetric
particles.  The lightest supersymmetric particle is essentially
always the $\cnone$ neutralino.  Next in line are the (roughly
degenerate) $\cntwo$
and $\cpmone$ neutralino and chargino states.  All three particles can be
quite light for moderate gluino mass. Indeed, for the standard
assumption of universal soft-supersymmetry-breaking
gaugino masses at the unification scale,
their masses scale with $\mgl$: very roughly
$\mcnone\sim {\mgl/ 8}\div{\mgl/ 6}$,
while $\mcpmone\sim\mcnone\sim {\mgl/ 4}\div{\mgl/ 3}$.
Since many models
allow $\mgl$ values as low as $200\gev$, and certainly $\mgl\sim 400\gev$
is possible for essentially all models, we see that
it is not at all unlikely to have $\mcnone\sim 20\div 50\gev$ (the
current experimental lower bound is $\sim 20\gev$ \cite{pdg}) and
$\mcntwo\sim\mcpmone\sim 45\div 100\gev$ (the current experimental
lower bounds being $\sim 45\gev$ \cite{pdg}).  In addition,
in the many models for which the soft scalar supersymmetry-breaking mass
terms are smaller than the soft gaugino supersymmetry-breaking
masses the sleptons are also quite light. In fact, in several
of the classes of model explored in Refs.~\cite{gunpois,bgkp} the sleptons
can be as light as or lighter than the $\cntwo$ and $\cpmone$.

Thus, we must consider carefully the impact on SUSY decay modes
of the MSSM Higgs upon LHC discovery prospects.
First, consider the $\hl$.  Generally speaking the only
SUSY decay modes that can be important for the $\hl$ are
$\hl\rta \cnone\cnone$ and $\hl\rta \snu\snu$; the first channel
is guaranteed to be invisible, and the second channel is highly
likely to be invisible.  Thus, we must ask whether or
not detection of an invisibly decaying $\hl$ at the LHC is possible.
Two modes have been explored:  $t\anti t \hl$ production
\cite{guninv} and $W\hl$ production \cite{kaneinv,royinv}.
In both cases, a lepton from $W\rta \ell \nu$ decay is used as a trigger,
and large missing transverse energy, $\etmiss$, is demanded.
Also useful is the effective mass constructed from the missing transverse
energy and the observed lepton, $\mmissl$.

In the case of $t\anti t \h$ (where $\h$ is any invisibly decaying
Higgs), one $t$ is required to decay leptonically ($t\rta Wb\rta l\nu b$)
and the other hadronically ($t\rta Wb\rta jjb$). One $b$-quark
tag is demanded (using $\ebtag,\emistag=30\%,1\%$
as always). Cuts requiring $M_{jj}\sim \mw$ (where
neither $j$ is the tagged $b$) and $M_{jjb}\sim\mt$ (where the $b$
is the tagged $b$-quark) are imposed. A large value of $\mmissl$
is also required.  Backgrounds include the irreducible process
$t\anti t Z$ (with $Z\rta \nu\anti \nu$) and reducible
backgrounds from $t\anti t (g)$-related tails deriving from
decays containing neutrinos.
The result, assuming SM-like $t\anti t \h$ coupling strength and
$BR(\h\rta {\rm invisible})\sim 1$, is that
for $\mt$ in the $\sim 140\div 180\gev$ range excess events
from $t\anti t \h$ production will be detectable above background
at the $\nsd=5$ level
for masses of $\mh=60,100,140,200$ in 0.2,0.3,0.5,1.6 nominal
$100\fbi$ LHC years, respectively.  Since the $\hl$ generally
has SM-like $t\anti t$ coupling for $\mha\gsim 1.5\mz$,
and since it almost certainly has mass below 140 GeV, we see
that an invisibly decaying $\hl$ will be detectable at the LHC.
It should be noted that this channel does not depend upon the
$WW\h$ coupling, and so is also potentially useful for the $\hh$
and $\ha$.

The $W\h$ detection channel yields somewhat similar results
assuming that the $WW\h$ coupling is of SM strength.
The $\mh$ mass reach of this channel is not quite as great.
However, $b$-tagging is not required, a possible advantage
in the high-luminosity LHC environment. Of course,
the $W\h$ mode is not useful for the $\ha$ or an $\hh$
with suppressed $WW\hh$ coupling.

Turning now to the $\ha$, $\hh$ and $\hp$, when heavy
they can all decay to a rather
wide selection of SUSY modes in models with light inos and/or sleptons.
When allowed, such modes will have significant branching ratios,
and are quite likely to be dominant. A brief outline
of the effect upon the previously discussed detection channels
is the following. At large $\mt\sim 200\gev$ and moderate $\tanb$,
the $\hh\rta 4\ell$ mode is an important component of the no-lose
theorem (crucial, if $b$-decay modes cannot be employed).  The
branching ratio for this mode is rapidly reduced as SUSY modes
become allowed (since the $\hh ZZ$ coupling is small compared to SM strength
for higher $\mhh$ values), and this signal would be largely lost.
At large $\tanb$, the $\tau^+\tau^-$ and
$b$-tagging modes are of primary concern.
Since the $b\anti b$ and $\tau^+\tau^-$ couplings
of the $\ha$ and $\hh$ are
enhanced at large $\tanb$, the $\ha,\hh\rta b\anti b,\tau^+\tau^-$
branching ratios can remain significant even in the presence
of SUSY decays, but would still
typically decrease by a factor of $2\div 3$ for the $\tanb\sim 15\div 6$
region of the discovery boundaries. Since the $b\anti b\ha$
and $b\anti b \hh$
associated production rates are roughly proportional to $\tan^2\beta$,
the boundaries would move to higher $\tanb$ by a factor
given roughly by $[BR(no~SUSY)/BR(with~SUSY)]^{1/2}$.
Finally, it is obvious that SUSY decays would make
the already difficult task of detecting the $\hp$ even
more challenging.

The question, that remains largely unanswered, is whether
or not the SUSY-particle-pair decays of a heavy Higgs
can be detected in the background from SUSY-particle-pair
continuum production. The answer will be very model dependent.
One mode that might yield a useful signal
is $\hh,\ha\rta \cntwo\cntwo$.
One would look for an excess of $4\ell$ events,
coming from the di-lepton decays of the $\cntwo$'s,
above the background from continuum $\cntwo\cntwo$ production ---
SM backgrounds are small.
This was explored in Ref.~\cite{baercntwo}. They found that
under fairly optimal conditions detection of the $\ha$ would
be possible.  However, should the $\cntwo$ decay primarily
to $\nu\snu$ (as occurs in many of the models explored
in Ref.~\cite{bgkp}) the $4\ell$ signal would be greatly reduced.
The corresponding mode for the $\hp$ is $\hp\rta \cntwo\cpone$.
However, here one would be focusing on tri-lepton events,
for which SM backgrounds are non-zero and discrimination against
continuum $\cntwo\cpone$ production would be more difficult.
Much work remains to be done before we can establish the
degree to which SUSY decay modes of a heavy MSSM Higgs can
be detected at the LHC.

\subsection{Prospects at the Di-Tevatron}

As summarized earlier, $\hsm$ detection at the Di-Tevatron
will be possible in the $W\hsm\rta \ell b\anti b X$ mode
for $\mhsm\lsim 100\div 110\gev$ (using conservative
cuts and resolutions).  This is not terribly different, perhaps
even a bit better, than the $\mhsm$ mass reach for this mode
at the LHC.  Thus, it is no surprise that Ref.~\cite{stangeetalwhsm}
finds that $W\hl$ detection will be possible at the Di-Tevatron
for a portion of parameter space that is comparable to that
discussed above in the case of the LHC.  Indeed, $W\hl$
detection at the Di-Tevatron may be possible for some section
of the $\mt=150\gev,\mstop=1\tev$ hole with only $L=50\fbi$;
and for $L=100\fbi$, $W\hl$ detection would be possible for all $\mha$
values that are in the hole or above.

This coverage would increase substantially if $\mstop$
is significantly lighter.
As already noted, in some unification contexts relatively modest $\mstop$
values are predicted (\eg\ $\mstop~\lsim 300\gev$) for much
of the allowed parameter space. Thus, even though $\mha\gsim2\mz$
in these same models, one finds
$\mhl\lsim 115\gev$ for most of parameter space if $\mt\sim170\gev$.
In this case, the Di-Tevatron could provide
a possibly crucial extension to the LEP-II discovery window, especially
if the latter is operated at a center of mass energy below 200 GeV.

Of course, as earlier noted,
the $\hl$ can have a large branching ratio to invisible
modes, in which case techniques based on $W\hl\rta \ell+\etmiss+X$
must be employed for discovery.  It seems likely that
the LHC techniques of Refs.~\cite{kaneinv,royinv} could
be employed to isolate the signal at the Di-Tevatron.

As for the $\ha$ and $\hh$,  once $\mha\gsim 1.5\mz$
their couplings to $WW$ and $ZZ$ are suppressed,
and their detection will be very difficult.
We have seen that even at the LHC
our ability to detect them is far from certain.
Detection of the $\hpm$ would be even more difficult unless
it happens that $t\rta \hp b$ is kinematically allowed,
and $\tanb$ is not in the small-$BR(t\rta \hp b)$-region
described earlier around $\tanb\sim 6$.

\section{MSSM HIGGS BOSONS WITH UNIFICATION}

One of the principle problems with the preceding discussion
is the large uncertainty for many of the MSSM parameters,
especially the SUSY partner masses.  The success of
gauge-coupling unification in the context of the MSSM
strongly suggests that one should also consider relatively
simple and universal boundary conditions for the soft-supersymmetry-breaking
parameters at the unification scale, $\mgut$.  Normally, a common
value is adopted at $\mgut$ for each of three categories of soft terms:
the squark, slepton and Higgs scalar masses, $m^0$;
the gaugino masses, $M^0$; and the soft-trilinear scalar Yukawa
couplings, $A^0$.

Specific supergravity/string
scenarios suggest definite relations among $m^0$, $M^0$ and $A^0$
\cite{bgkp}.
A fairly representative example of one class of models
is that provided by the so-called
`no-scale' or minimal supergravity choice of $m^0=A^0=0$.
Many other well-motivated model classes yield results similar to those we
will describe for the minimal supergravity model. Among these are
the dilaton and dilaton-like models,
where the dilaton provides the main source of supersymmetry
breaking. For these models, $m^0$ and $A^0$ are non-zero but scale with $M^0$
with a proportionality constant smaller than unity.  The important
common features of these models that influence Higgs phenomenology are three:
a) the low-energy physical masses of the stop squarks are substantially
smaller than the
gluino mass, $\mgl$, so that radiative corrections to the Higgs sector
are not that large for a big chunk of parameter space;
b) slepton masses are often comparable to or lower than
the masses of the $\cntwo$ and $\cpmone$,
implying not only that the $\snu$ can decay invisibly but also that
$\snu\snu$ pairs can be a dominant decay mode for the $\hl$;
and
c) the lower bound on the $\mha$ mass from combining experimental
constraints with the requirement of a correct EWSB potential minimum
is substantial.
Of course, there is an important class of
models with $M^0<m^0$, and such models can have quite different
Higgs phenomenology in the above respects.

Given the value of the universal soft gaugino mass $M^0$,
and $A^0$ and $m^0$ values that are determined (either to
be zero, as in the model we shall specifically discuss,
or as numbers proportional to $M^0$), only three parameters
are required in order to fully specify the theory within the SUSY plus
renormalization group evolution context.
When considering the Higgs sector these are most conveniently chosen to be
$\tanb$, $\mha$ (which low-energy-scale parameter
can be though of as being traded for the $\mgut$-scale parameter $M^0$)
and one other condition or parameter.  A popular idea is
to demand that the $b$ and $\tau$ Yukawa couplings be equal at
the unification scale \cite{pokorski}, as required in models such as
flipped $SU(5)\times U(1)$.  With this final constraint imposed, a choice of
$\mha$ and $\tanb$ completely determines the theory, leaving us with
the $\mha-\tanb$ parameter space that we have been discussing.
This model has been explored in Ref.~\cite{gunpois}.

The highly constrained nature of
this model context, denoted by `Yukawa-Unified, Minimal Supergravity'
or YUMS, has many implications. First, it can happen
that a given choice of $\mha,\tanb$ is simply inconsistent with
correct electroweak
symmetry breaking (for instance, the residual electromagnetic $U(1)$
symmetry could be spontaneously broken or color symmetry could be broken)
or with experimental constraints (\eg,
some SUSY particle masses could be inconsistent with lower bounds
from LEP). In the YUMS model, these constraints limit
the $\mha-\tanb$ parameter
space to the allowed domain illustrated in Fig.~\ref{yumsfig}.
Note that $\mha\gsim 200\gev$ for most of the allowed parameter
space, while $\tanb$ lies between about 1.7 and 10, with lower
values being required for lower $\mt$ (see the right-hand axis
of the graph for $\mt$ as a function of $\tanb$).

%ffffffffffffffffffffffffffffffffffffffffffffffffffffff
\vspace{.5cm}
\begin{figure}[htbp]
\begin{center}
%\mbox{
%\epsfysize=7.5cm
%\epsffile{zeuthen_yums_smallsize.ps}
%}
%\mbox{
\leavevmode
{\psfig{figure=zeuthen_yums_smallsize.ps,width=3in,clip=}}
%     }
\end{center}
\caption{Higgs discovery in the YUMS model at LEP-II and LHC.}
\label{yumsfig}
\end{figure}
%------------------------------------------------------------

It is interesting to observe that the CDF $\mt$ (`pole') mass value
of $\mt(pole)\sim 174\gev$ corresponds to a running $\mt(\mt)$ mass
of about 167 GeV, which in turn requires that $\tanb$ lie below
2 in this model.  This is a general correlation that arises
in the Yukawa-unified MSSM context \cite{pokorski}.
Higgs sector implications of such a small $\tanb$ value will
become apparent below, but we mention the most obvious one now: $\mhl$ is
quite moderate in size, even after radiative corrections, for low $\tanb$.

The most crucial point compared to our previous general
MSSM Higgs phenomenology discussions
is that specification of $\mha$ and $\tanb$ determines {\it all}
masses, parameters, branching ratios and so forth, for all sectors
of the theory.  For example, since the masses of the top quark
and of the stop squarks will be functions
of $\mha$ and $\tanb$, we can compute $\mhl$ (including radiative corrections)
as a function of $\mha$ and $\tanb$.  In the YUMS model,
the lightest stop has $\mstop\sim 400\gev$ for almost all of parameter space.
Thus, we find that $\mhl\lsim 105\gev$, \ie\ $\mhl$ is such that $Z\hl$
will be a useful detection mode at LEP-200, for all but the small
portion of parameter space outlined by dots in Fig.~\ref{yumsfig}.
Similarly optimistic results arise for the other closely related
models explored in Ref.~\cite{bgkp} --- in all these models the $\hl$ is
light enough that $Z\hl$ detection will be possible at LEP-200
for much of allowed parameter space.
But, we must again stress that in other types of models, namely those
having boundary conditions with $M^0<m^0$,
$\mstop$ can be very large and the $\hl$ much heavier.

The conclusion that LEP-II will find the $\hl$ in the YUMS and related
models is, as already indicated
by the unconstrained-model discussion given earlier,
very sensitive to the energy of LEP-II. If LEP-II operates
at $\sqrt s\sim 176\gev$, then $\hl$ masses up to only about $80\gev$
can be probed, and $Z\hl$ production will only be visible
for less than half of the parameter space of Fig.~\ref{yumsfig}.
However the viable region would include the small $\tanb$
values associated with the CDF top-quark mass estimate.
In this model, a Di-Tevatron, which we have seen can also probe
up to $\mhl\sim 100\div 110\gev$, would be a very useful machine
if LEP-II does not reach $\sqrt s\sim 200\gev$.

Of course, the decay modes of the $\hl$ determine the
final states that must be employed for its detection.
Since the masses of the lightest neutralino
($\cnone$) and generally-invisible sneutrino ($\snu$)
are determined as functions of $\mha$ and $\tanb$,
we can compute the relative importance
of invisible decays of the $\hl$. The region where the branching
ratio for such decays
is sufficiently large that invisible detection modes must be employed
at the LHC is also shown in Fig.~\ref{yumsfig} (the region to the left of the
internal solid line).  In this same region, $Z\hl$ detection at LEP-II
requires using missing-mass reconstruction techniques.
The invisible decays occur at small $\mha$ since, in the constrained
YUMS model context, small $\mha$ is associated with small $M^0$ and thence
with small masses for the lightest neutralino and chargino.
Note that the modest $\mhl$ values predicted for modest $\tanb$ do not
prevent invisible decays from being important.

Finally, Fig.~\ref{yumsfig} also indicates the regions where the
earlier-described $\gam\gam$ inclusive, $\ell\gam\gam$ associated
production, and $t\anti t \hl\rta \ell 3b X$ (as well as $W\hl\rta \ell 2bX$)
modes will allow $\hl$ detection at the LHC (indicated by the regions
to the right of the dotdash, dotdotdash, and dashed contours, respectively).
Note the sharp transition between the region for which these modes are
viable and the region for which the invisible detection modes
must be employed at the LHC. To be able to detect
the $\hl$ for all of parameter space,
the LHC detectors must be certain of having the capability to
explore both classes of modes.  Detection in the invisible modes requires
a high degree of detector hermeticity.

An important feature of this and related models is
that {\it only the $\hl$ will be detectable} at hadron colliders
and LEP-II. The modest value for $\tanb$ alone is sufficient to
bring into question detection of
the $\ha$ and $\hh$ in either the $\tau^+\tau^- b\anti b$ or $4b$ modes,
and is neither small enough nor big enough
for $\hpm$ detection in the $t\anti t b\anti b$
final state, even if we assume no depletion of the required decays
by SUSY-particle-pair final states.  In fact, in the YUMS
model (and its cousins discussed in Ref.~\cite{bgkp}) the decays of the $\ha$,
$\hh$ and $\hpm$ are dominated by a plethora of SUSY final states, making
detection still more problematical.  Such decays even make $\ha$, $\hh$
and $\hp$ detection at an NLC with adequate energy for $\hp\hm$
and $\ha\hh$ pair production (\ie\ an NLC
having $\sqrt s \gsim 2.5 \mha$) less than certain.

Whether or not the YUMS model is the correct theory, it illustrates
the types of special situations and complexities that can easily
arise if the MSSM is part of a larger unified theory.
The LHC experimental collaborations must develop
sufficiently flexible detectors
that sensitivity to all the above-described detection modes is
implemented.

\section{CONCLUSIONS}

In conclusion, it is clear that the roles and relative importance of
LEP-II, the LHC, an NLC  and a possible Di-Tevatron are dependent
on many factors.  On the experimental side,
for the hadron colliders, the achievable
luminosity and ability to $b$-tag at high luminosity play
a crucial role. For LEP-II and NLC, the actual center-of-mass
energy is crucial in establishing expected mass reach.
On the theoretical side, there is enormous model dependence.

At one extreme we can assume that the SM is correct, and
assign equal probability
to any $\hsm$ mass below the perturbative limit of $600\div 800$ GeV.  Then,
if the NLC is assumed to have $\sqrt s \sim 500\gev$,
only the LHC guarantees Higgs detection. But complete verification
of the properties of the $\hsm$ (\eg\ branching ratios and couplings)
at the LHC will be very difficult.  For instance, for $\mhsm\lsim 120\gev$ the
primary $\hsm$ decay mode is to $b\anti b$,
but $\hsm$ observation in this mode
requires highly efficient and pure $b$-tagging
in a high-luminosity environment. In contrast, if the NLC
energy is sufficient to produce the $\hsm$, reasonably accurate
determination of its decay branching fractions and basic couplings will be
possible \cite{hildreth,gunhawaii,janothawaii}.

At the other extreme, a unified MSSM model, such as the YUMS model,
could be correct, predicting that the $\hl$ is very likely to be
within the reach of LEP-200 (but not necessarily LEP-176).
Although the $\hl$ would be seen at
the LHC, and probably in several modes that would provide
some information on branching ratios and couplings (though not as
much as would be available at LEP-200), detection of the other (heavier) MSSM
Higgs bosons would probably require an NLC with center-of-mass
energy substantially above 500 GeV.  Even there, the dominance of SUSY
decay modes of the heavier Higgs could make their detection quite tricky.
Meanwhile, should LEP-II not reach 200 GeV or above, a Di-Tevatron
upgrade would provide a significant addition to coverage of the
mass range $\mhl\lsim 115\gev$ predicted in a YUMS-like model.
More generally, unified MSSM models will fall in between these extreme
situations, and LEP-II, the LHC and the NLC
would play highly complementary roles,
while the Di-Tevatron would provide some $\hl$ discovery potential
beyond LEP-II limits.

Alternative Higgs sector models provide a further broadening
of the spectrum of possibilities.  A full arsenal of
complementary accelerators would certainly be the most optimal
approach to exploring the Higgs sector and electroweak symmetry
breaking, and would in any case be needed to guarantee the ability
to discover and explore the other new particles and phenomena
that in general (\eg\ in supersymmetric and technicolor models)
accompany the EWSB mechanism.

\section{ACKNOWLEDGEMENTS}

I would like to thank DESY-Zeuthen for their kind hospitality
during my visit, and for a most enjoyable and informative conference.
I am grateful for the support and hospitality of the Stanford
Linear Accelerator during the course of preparing this report.
The implicit contributions of all my collaborators to the
summary presented here are gratefully acknowledged.
This work was supported in part by the U.S. Department of Energy.

\end{document}